# Bike Renting Data Analysis: The Case of Dublin City

Thanh Thoa Pham Thi[*], Joe Timoney, Shyram Ravichandran, Peter Mooney, Adam Winstanley

Department of Computer Science, Maynooth University

**Summary**

Public bike renting is more and more popular in cities to incentivise a reduction in car journeys and to boost the use of green transportation alternatives. One of the challenges of this application is to effectively plan the resources usage. This paper presents some analysis of Dublin bike renting scheme based on statistics and data mining. It provides available bike patterns at the most interesting bike stations, that is, the busiest and the quietest stations. Consistency checking with new data reinforces confidence in the patterns obtained. Identifying available bike patterns helps to better address user needs such as organising the rebalancing of the bike numbers between stations in advance of demand.

**KEYWORDS:** bike renting, data analysis, clustering, data visualisation, Open Street Maps

## 1. Introduction

Public bike sharing has been implemented in over 600 cities today. Dublin-bikes is considered to be one of the most successful bike sharing schemes in the world. It has more than 64,000 subscribers and 16.3 million journeys have been taken since the scheme started in 2009 (DublinBikes, 2009). Recently, the inclusion of access to Dublin-bikes using the Leap Card, a public transport prepaid smart card, has facilitated wider usage of the scheme (DublinBikes, 2009).

Bike renting data analysis for the big cities have been increasingly carried out. (Schuijbroek et al., 2016) has grouped the existing work into four research streams, in which we are interested in (i) *demand analysis* which helps to predict the future demand and to identify the impact factors for managerial decision making, (ii) *service level analysis* which focuses on the available bikes and docks (Gast et al., 2015), and (iii) *rebalancing operations* (Vogel and Mattfeld, 2010), (Schuijbroek et al., 2016).

However, little analysis has been done on Dublin bike data. In 2010, (Mooney at al. 2010) presented some preliminary analysis on Dublin bike data where there were only 40 stations and 450 bikes. Since then, the scheme has been broadened significantly, with 101 stations and around 1580 bikes.

In this paper, we perform some demand analysis and service level analysis on Dublin bike data using our own analysis process. Firstly some statistical analysis on check-in and check-out bike is described to identify the most interesting stations: these are the busiest and the quietest stations. We then explore these more deeply by discovering specific available bike patterns over time using a clustering technique. To validate the patterns discovered, some tests are carried out with new data. The available bike patterns depend on the station location and time. They also corroborate with intuitive knowledge regarding the surrounding facilities. This analysis helps to better understand user behaviours which leads to insights that facilitates planning, such as the rebalancing the bikes across stations in advance of demand.

## 2. Data collection

The data was gathered from the 13th September 2016 until the middle of December 2016 using the

---
[*] thoa.pham@nuim.ie

JCDecaux API (JCDecaux, 2016). Data is collected every 10 minutes from 12am to 11.59pm every day in that period. The data collected is in JSON format and is then converted to be stored as a .csv file. Over 24000 records per day were logged and each record contains the information as presented in Table 1.

**Table 1. Structure of collected data**

| Column | Description |
|---|---|
| 1 | Station number |
| 2 | Station Name |
| 3 | Position/Latitude |
| 4 | Position/Longitude |
| 5 | Total stands |
| 6 | Available stands |
| 7 | Available bikes |
| 8 | Last update |

### 3. Exploration analysis

Our first analysis is to calculate the level of traffic at each station on a daily basis, which corresponds to the total of the check-ins (dropping-in bike) number and the check-outs (taking-out bike) number per day during the period of data collection. The traffic level represents the activity degree of a station. It is very intuitive to visualise it using a map-based display. In Figure 1, the map shows the location of each station with an orange circle and the bigger the circle, the busier the station. The day of interest can be changed using a slider at the top of the map. It was observed that for some stations there is stability in the traffic levels on different days, while for others there are some remarkable differences.

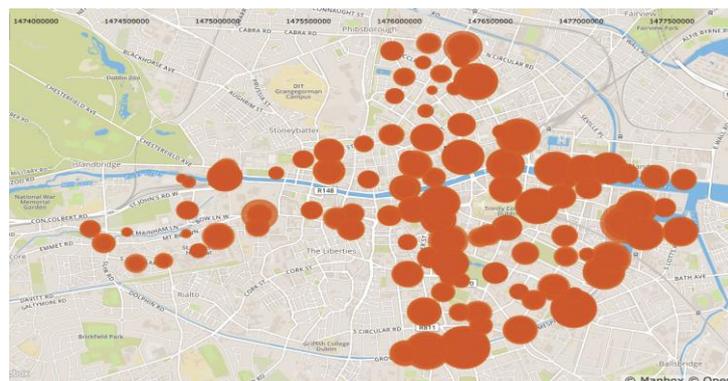

**Figure 1**. A map of the Dublin-bike stations and their degree of activity illustrated by the width of orange circles

This can be explained based on the station location. For instance, Charlemont Place and Custom House stations were found to be very busy during weekdays and weekends, while Heuston station is busy on weekdays but very quiet at weekends. Charlemont place and Custom house are located between busy residential and business districts. Charlemont place is at the entrance to Ranelagh, a very popular area for renting. Custom House is just beside the IFSC (Irish Financial Services Centre), a mixed use area that is populated primarily with young professionals. The stops are approximately 2km and 1km respectively from the city centre, with many road crossings that can delay a pedestrian. Their locations mean that they could easily be busy all week with people cycling for work and for pleasure. In contrast, Heuston station is beside one of the two national railway stations, connecting with the south, south west and the west. It supports regular commuter rail traffic from the west of the city. Additionally, just at its entrance is a major stop for the Luas tram to the western suburbs. The city centre is located at a distance of 2km. To walk to the centre from Heuston station is time consuming and thus other forms of transport are preferable. Many people would pass through Heuston station during the week on their way to work and would be located within a radius of the city centre. Cycling would be an attractive option for two reasons. Firstly, public transport options of the Luas tram and buses can be slow because there are a number of stops on the journey to the centre. Secondly, if someone wants to get to a point away from the main road into the centre it is difficult as all bus routes radiate outwards from the centre rather than operate crossways trough the city. Thus, cycling is an efficient option. At weekends the people going through Heuston station are not working commuters. They would more likely be carrying baggage thus inhibiting the use of a Dublin-bike.

It is worthwhile to further identify any patterns in term of traffic changes over time during a day of the stations. This is explored in the next analysis.

## 4. Pattern analysis and discussions

The analysis at this phase takes into account the number of available bikes every 10 minutes, every day at each station during the observation period. Firstly the data of the busiest station, Charlemont Place, was investigated. The K-means algorithm was used to group the bikes-available data of this station into 4 clusters. Figure 2 shows the representative element of 4 clusters, which is the centroid, as it changes for each cluster over the times of the day. The variation in the value of the centroid is greater for some clusters than others and illustrates a strong time-dependence in the case of three of the clusters. Further investigation found that the clusters are significantly influenced by the days of the week, as shown in Table 1. Cluster 0 includes 6 items on Saturday, this means that the data within Cluster 0 is coming from data for Saturdays exclusively. Cluster 1 has data from 4 Saturdays, 9 Sundays and 1 Monday. Actually that particular Monday is a Bank holiday in Ireland (30$^{th}$ of October), and thus is of similar character to a weekend day. Clusters 2 and 3 have data from week days primarily.

An interpretation can be made from examining both Figure 2 and Table 2 together. Cluster 0 indicates that on Saturdays, the number of available bikes peaks at around 9.30AM and then decreases quickly until 3pm and then continues decreases at a slower rate until midnight. Cluster 1 captures the pattern of activity on Sundays, in which the number of available bikes is low from mid-night, it from 9.30 AM to 11.30 AM, and then lightly decreases in the vicinity of mid-day and then mostly keeps increasing at a gentle rate until midnight. Clusters 2 and 3 mostly represent the number of available bikes during weekdays. The data on Monday, Tuesday and Wednesday is grouped perfectly into Cluster 2, and Thursday and Friday fall into Cluster 3. The patterns of these 2 clusters are very similar until around 5.10pm, after which the difference is noticeable. This big difference can be hypothesized as follows: at the beginning of the week people may not go out much in the night time after work so they would not use the bikes and the number of available bikes is high after 5pm. However, closer to the weekend on Thursday and Friday people tend to go out during the night time and the bikes are a popular choice for getting around and having a drink. The graph shows a similar trend for Sunday night, which means that the number of available bikes at this station is much lower than on Monday, Tuesday and Wednesday nights.

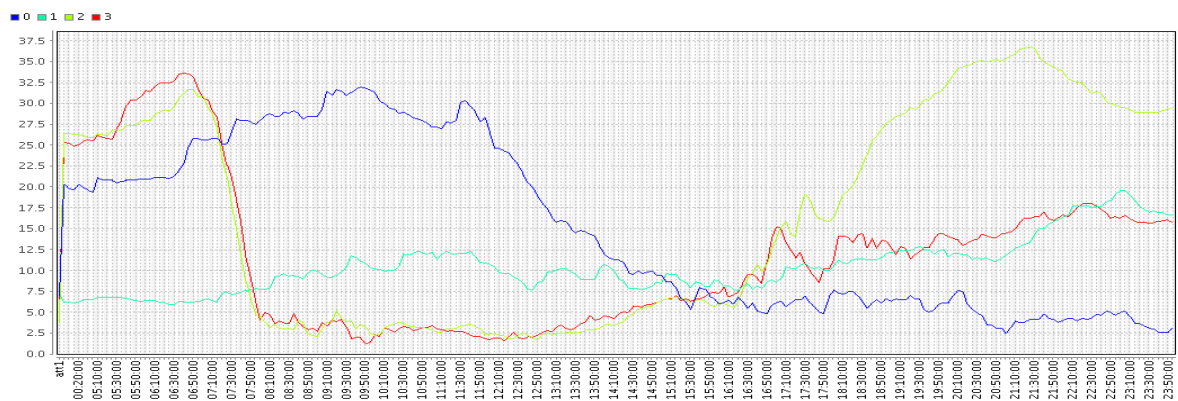

**Figure 2.** Centroids of 4 clusters for Charlemont Place station shown over the timeline of a day
Cluster 0 (blue line), Cluster 1 (cyan line), Cluster 2 (green line), Cluster 3 (red line),

**Table 2**. Details of each cluster at Charlemont station

| Cluster | Mon | Tue | Wed | Thu | Fri | Sat | Sun |
|---|---|---|---|---|---|---|---|
| Cluster 0 | | | | | | 6 | |
| Cluster 1 | 1 | | | | | 4 | 9 |
| Cluster 2 | 8 | 10 | 7 | 6 | 4 | 0 | 1 |
| Cluster 3 | | 1 | 4 | 5 | 7 | | |

We also examined data for the quietest station, Heuston. For this station, two interesting clusters are found, Cluster 0 and Cluster 1 which correspond to weekend data and weekday data respectively (Fig. 3). The Monday that belongs to Cluster 0 in Table 3 is actually a Public holiday in Ireland.

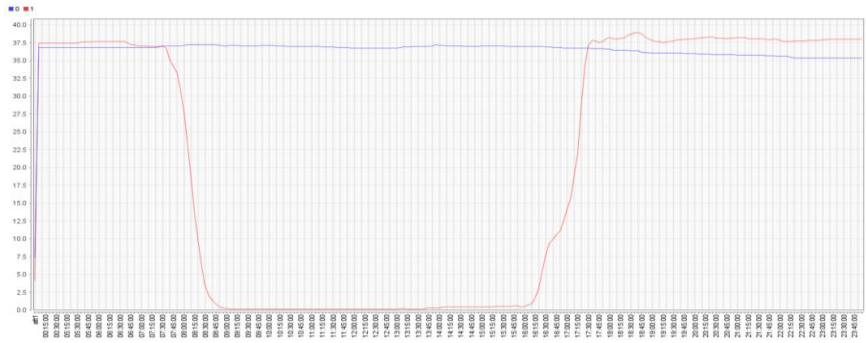

**Figure 3**. Centroids of 2 clusters for Heuston station over the timeline of a day

**Table 3**. Details of each cluster at Heuston station

| Cluster | Mon | Tue | Wed | Thu | Fri | Sat | Sun |
|---|---|---|---|---|---|---|---|
| C0 | 1 | | | | | 10 | 10 |
| C1 | 8 | 11 | 11 | 11 | 11 | | |

The clusters provide the following interpretation: during weekends the user are not working people, so with their luggage it is not convenient to use the bikes. The number of available bikes is very stable over both days. Meanwhile there is a huge difference in the number of available bikes between the business hours and out of business hours for weekdays. For instance, during weekdays people starts taking bikes here from 7.30 am, the number of available bikes is 0 during business hours, then it increases to the maximum from 5.30pm and keeps the same number to until the morning.

To validate the analysis results, we have tested the patterns observed against new data collected in the first 2 weeks of December. The comparison of the new data with the clusters identified at the stations demonstrates consistency. For instance, items on a weekday (or weekend) in December fit with the cluster corresponding to weekday (or weekend) as shown graphically in Figures 4 and 5.

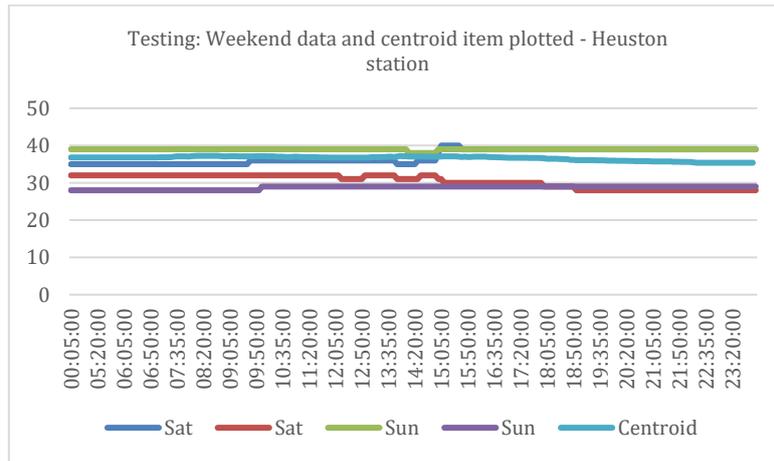

**Figure 4.** Testing new weekend data for Heuston station

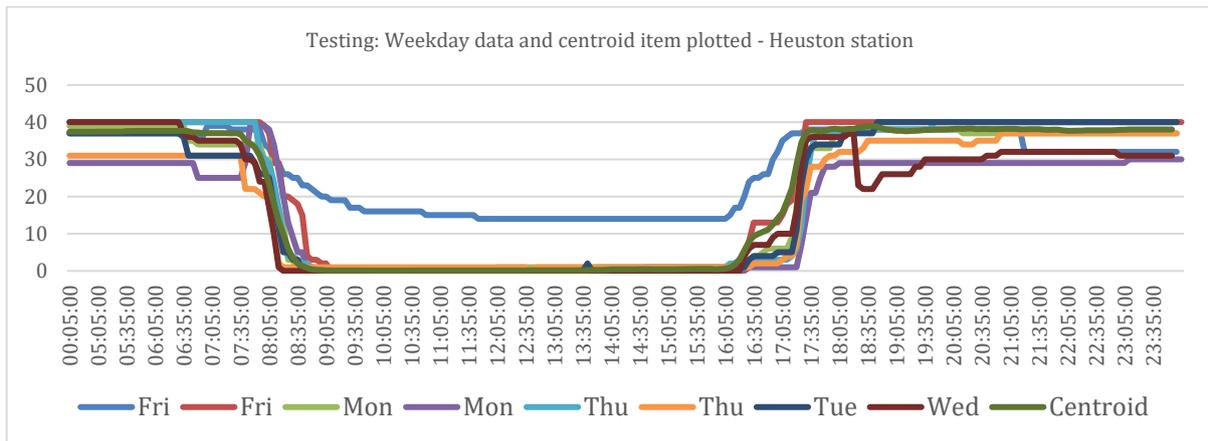

**Figure 5.** Consistency check for new weekday data at Heuston station

## 5. Conclusion

Analysis of Dublin bike data can be utilized to provide insights into bike usage patterns. In this paper, we have performed clustering analysis and identified interesting clusters at the busiest and quietest bike stations. Examining the number of available bikes every 10 minutes, the results showed that at the stations analysed there can be significant differences in bike usage at weekends and weekdays, during business hours and after work. The different patterns observed from the results can be explained based on geographical knowledge regarding the locations. Testing with new data suggests pattern consistency. The results from this phase can be used as the starting point for developing a more rigorous prediction analysis on the number of available bikes at any given station.
In addition, we are going to combine more factors into bike data analysis such as weather conditions to get a complete insights of the user behaviors.

**Biography**

Thanh Thoa Pham Thi, Joe Timoney and Peter Mooney are lecturers in the Computer Science Department, Maynooth University. Adam Winstanley is Professor of the same Department. Shyram Ravichandran is a Master student in Computer Science of the Maynooth University.